\documentclass{wscpaperproc}
\usepackage{latexsym}
\usepackage{graphicx}
\usepackage{mathptmx}

\usepackage{amsmath}
\usepackage{amsfonts}
\usepackage{amssymb}
\usepackage{amsbsy}
\usepackage{amsthm}
\usepackage[pdftex,colorlinks=true,urlcolor=blue,citecolor=black,
anchorcolor=black,linkcolor=black]{hyperref}

\usepackage{dcolumn}
\newcolumntype{d}[1]{D{.}{.}{#1}} % column in mathmode and decimal point aligned

\newtheoremstyle{wsc}% hnamei
{3pt}% hSpace abovei
{3pt}% hSpace belowi
{}% hBody fonti
{}% hIndent amounti1
{\bf}% hTheorem head fontbf
{}% hPunctuation after theorem headi
{.5em}% hSpace after theorem headi2
{}% hTheorem head spec (can be left empty, meaning `normal')i

\theoremstyle{wsc}

\begin{document}

%***************************************************************************
% AUTHOR: AUTHOR NAMES GO HERE
% FORMAT AUTHORS NAMES Like: Author1, Author2 and Author3 (last names)
%
%		You need to change the author listing below!
%               Please list ALL authors using last name only, separate by a comma except
%               for the last author, separate with "and"
%

% setting up general page style
\pagestyle{fancyplain}

% setting up page style of first page
\thispagestyle{plain}
\firstPageHead{}

% setting up running header (authors) of subsequent pages
\chead{\fancyplain{}{\itshape Uhlig, Hillmann, and Rose}}

% setting up seperation parameters
%\headsep=72pt
\rhead{}
\cfoot{}
\renewcommand{\headrulewidth}{0pt} % (renewcommand needed in fancyhdr to remove top decorative line)
%\headrulewidth=0pt  % ("setlength" needed in fancyheading to remove top decorative line)

%%%%%%%%%%%%%%%%%%%%%%%%%%%%%%%%%%%%%%%%%%%%%%%%%%%%%%%%%%%%%%%%%%%%%%%%%%%%%%
%                                                                            %
%     THESE COMMANDS ARE REQUIRED TO WORK WITH WSC.BST TO MAKE BIBLIO     %
%                                                                            %
%%%%%%%%%%%%%%%%%%%%%%%%%%%%%%%%%%%%%%%%%%%%%%%%%%%%%%%%%%%%%%%%%%%%%%%%%%%%%%
\makeatletter
\let\@internalcite\cite
\def\cite{\def\@citeseppen{-1000}%
    \def\@cite##1##2{(##1\if@tempswa , ##2\fi)}%
    \def\citeauthoryear##1##2##3{##1 ##3}\@internalcite}
\def\citeNP{\def\@citeseppen{-1000}%
    \def\@cite##1##2{##1\if@tempswa , ##2\fi}%
    \def\citeauthoryear##1##2##3{##1 ##3}\@internalcite}
\def\citeN{\def\@citeseppen{-1000}%
%  Pierre L'Ecuyer's fix for multiple cite bug
%  Added by Paul J Sanchez on 4 October 2001
%   \def\@cite##1##2{##1\if@tempswa , ##2)\else{)}\fi}%
%   \def\citeauthoryear##1##2##3{##1 (##3}\@citedata}
    \def\@cite##1##2{##1\if@tempswa, ##2)\else{}\fi}%
    \def\citeauthoryear##1##2##3{##1 (##3)}\@citedata}
\def\citeA{\def\@citeseppen{-1000}%
    \def\@cite##1##2{(##1\if@tempswa , ##2\fi)}%
    \def\citeauthoryear##1##2##3{##1}\@internalcite}
\def\citeANP{\def\@citeseppen{-1000}%
    \def\@cite##1##2{##1\if@tempswa , ##2\fi}%
    \def\citeauthoryear##1##2##3{##1}\@internalcite}
\def\shortcite{\def\@citeseppen{-1000}%
    \def\@cite##1##2{(##1\if@tempswa , ##2\fi)}%
    \def\citeauthoryear##1##2##3{##2 ##3}\@internalcite}
\def\shortciteNP{\def\@citeseppen{-1000}%
    \def\@cite##1##2{##1\if@tempswa , ##2\fi}%
    \def\citeauthoryear##1##2##3{##2 ##3}\@internalcite}
\def\shortciteN{\def\@citeseppen{-1000}%
%  Pierre L'Ecuyer's fix for multiple cite bug
%  Added by Paul J Sanchez on 2 September 2002
%  should have caught this last year...
%   \def\@cite##1##2{##1\if@tempswa , ##2)\else{)}\fi}%
%   \def\citeauthoryear##1##2##3{##2 (##3}\@citedata}
% Shane G. Henderson fix for extra right bracket at end of optional material June 8, 2005
%    \def\@cite##1##2{##1\if@tempswa, ##2)\else{}\fi}%
    \def\@cite##1##2{##1\if@tempswa, ##2\else{}\fi}%
    \def\citeauthoryear##1##2##3{##2 (##3)}\@citedata}
\def\shortciteA{\def\@citeseppen{-1000}%
    \def\@cite##1##2{(##1\if@tempswa , ##2\fi)}%
    \def\citeauthoryear##1##2##3{##2}\@internalcite}
\def\shortciteANP{\def\@citeseppen{-1000}%
    \def\@cite##1##2{##1\if@tempswa , ##2\fi}%
    \def\citeauthoryear##1##2##3{##2}\@internalcite}
\def\citeyear{\def\@citeseppen{-1000}%
    \def\@cite##1##2{(##1\if@tempswa , ##2\fi)}%
    \def\citeauthoryear##1##2##3{##3}\@citedata}
\def\citeyearNP{\def\@citeseppen{-1000}%
    \def\@cite##1##2{##1\if@tempswa , ##2\fi}%
    \def\citeauthoryear##1##2##3{##3}\@citedata}
%
% \@citedata and \@citedatax:
%
% Place commas in-between citations in the same \citeyear, \citeyearNP,
% \citeN, or \shortciteN command.
% Use something like \citeN{ref1,ref2,ref3} and \citeN{ref4} for a list.
%
\def\@citedata{%
    \@ifnextchar [{\@tempswatrue\@citedatax}%
                  {\@tempswafalse\@citedatax[]}%
}

\def\@citedatax[#1]#2{%
\if@filesw\immediate\write\@auxout{\string\citation{#2}}\fi%
  \def\@citea{}\@cite{\@for\@citeb:=#2\do%
    {\@citea\def\@citea{, }\@ifundefined% by Young
       {b@\@citeb}{{\bf ?}%
       \@warning{Citation `\@citeb' on page \thepage \space undefined}}%
{\csname b@\@citeb\endcsname}}}{#1}}%

% don't box citations, separate with ; and a space
% also, make the penalty between citations negative: a good place to break.
%
\def\@citex[#1]#2{%
\if@filesw\immediate\write\@auxout{\string\citation{#2}}\fi%
  \def\@citea{}\@cite{\@for\@citeb:=#2\do%
    {\@citea\def\@citea{, }\@ifundefined% by Young
       {b@\@citeb}{{\bf ?}%
       \@warning{Citation `\@citeb' on page \thepage \space undefined}}%
{\csname b@\@citeb\endcsname}}}{#1}}%

% (from apalike.sty)
% No labels in the bibliography.
%
\def\@biblabel#1{}
\makeatother

%\newlength{\bibhang}
%\setlength{\bibhang}{2em}

% Indent second and subsequent lines of bibliographic entries. Taken
% from openbib.sty: \newblock is set to {}.
% \renewcommand{\refname}{REFERENCES}

\newdimen\bibindent
\bibindent=0.0em
% SEC: was \def\thebibliography#1{\section*{\refname\@mkboth
% SEC: was   {\uppercase{\refname}}{\uppercase{\refname}}}\list
\def\thebibliography#1{\section*{\refname}\list
   {}{\settowidth\labelwidth{[#1]}
   \leftmargin\parindent
   \itemindent -\parindent
   \listparindent \itemindent
   \itemsep 0pt
   \parsep 0pt}
   \def\newblock{}
   \sloppy
   \sfcode`\.=1000\relax}

           % Set up BiBTeX macros

% needed to make the tex document look more like the word counterpart :-(
\setlength{\baselineskip}{12.7pt}

% AUTHOR: Enter the title, all letters in upper case
\title{EVALUATION OF THE GENERAL APPLICABILITY OF DRAGOON FOR THE $k$-CENTER PROBLEM}

% AUTHOR: Enter the authors of the article, see end of the example document for further examples
\author{Tobias Uhlig\\
Peter Hillmann\\
Oliver Rose\\ [12pt]
Department of Computer Science\\
Universit\"at der Bundeswehr M\"unchen\\
Werner-Heisenberg-Weg 39\\
Neubiberg, 85577, GERMANY
}

\maketitle

\section*{ABSTRACT}

The $k$-center problem is a fundamental problem we often face when considering 
complex service systems. Typical challenges include the placement of warehouses 
in logistics or positioning of servers for content delivery networks. We 
previously have proposed \emph{Dragoon} as an effective algorithm to approach 
the 
$k$-center problem. This paper evaluates \emph{Dragoon} with a focus on potential 
worst case behavior in comparison to other techniques. We use an evolutionary 
algorithm to generate instances of the $k$-center problem that are especially 
challenging for \emph{Dragoon}. Ultimately, our experiments confirm the 
previous 
good results of \emph{Dragoon}, however, we also can reliably find scenarios 
where it is clearly outperformed by other approaches.

\section{INTRODUCTION}

\begin{figure}[!b]
\centering
\begin{minipage}{.4\textwidth}
  \fbox{\includegraphics[width=\textwidth]{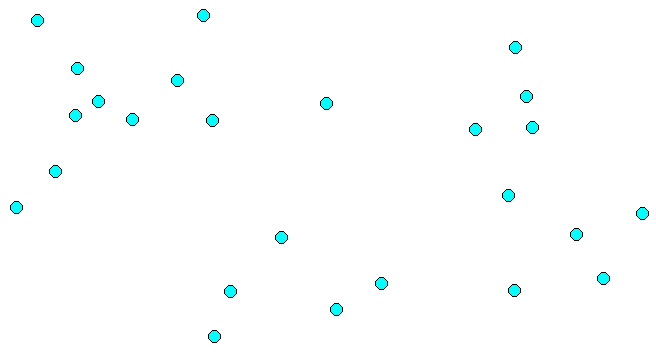}}
\end{minipage}%
\hfill
\begin{minipage}{.15\textwidth}
  $\xrightarrow{optimization} $
\end{minipage}%
\begin{minipage}{.4\textwidth}
  \fbox{\includegraphics[width=\textwidth]{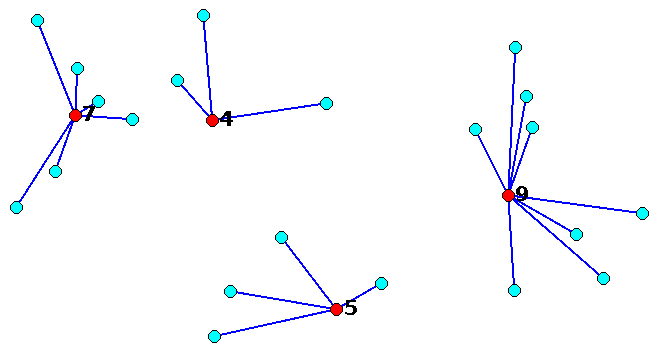}}
\end{minipage}
\caption{The $k$-center problem: for a set of customers the challenge is to 
determine a given number of central locations to provide a service to the 
customers. \label{fig:kcenter}}
\end{figure}

Imagine a manager that plans new warehouses to distribute goods to his customers. 
Consider a planner of content delivery networks who has to setup servers close 
to users to provide reliable access to information. Think of an emergency staff 
that needs to know where to place distribution centers to supply basic goods 
after a natural disaster. Complex service system like these and many other 
scenarios are instances of the $k$-center problem. Consequently, there is a big 
interest in solving these problems. Especially considering the last two examples 
we need reliable results in short time. Accordingly, an appropriate optimization 
approach needs to be robust and efficient. Figure \ref{fig:kcenter} shows an 
abstract instance of the $k$-center problem and a solution generated by the 
Dragoon algorithm. A general introduction to the $k$-center problem and respective 
optimization heuristics can be found in \citeN{Kleinberg2005}.

In this paper we discuss Dragoon in detail, an optimization approach that has 
previously been identified as an efficient optimization technique 
\shortcite{hillmann2015c}. We focus more on the robustness aspect of the 
optimization, by analyzing its worst case behavior in comparison to other 
techniques. To this end we employ an evolutionary algorithm to generate problem 
instances that are especially challenging for Dragoon. This idea is inspired by 
a paper from \shortciteNP{nguyen2014}. They used an evolutionary algorithm to 
generate images that would fool a deep neural network. The considered deep 
neural network was trained to recognized certain classes of images, however, the 
research group was able to generate images that were false positives. Similarly, 
we try to identify instances of the $k$-center problem were the explicit and 
implicit optimization assumptions of Dragoon are exploited, leading to low 
quality solutions.

Our paper is structured in the following way: In the next section we briefly 
introduce the $k$-center problem and its characteristics. Afterward, we present 
the optimization approaches we considered. In Sections \ref{experiment-setup} 
and \ref{experiment-results} we explain our experimental setup and provide the 
results of our tests. Finally, we discuss the results of our study and give a 
summary and short outlook.

\section{THE K-CENTER PROBLEM}

The $k$-center problem is a classic optimization problem, that is known to be 
NP-hard \cite{gonzales1985}. Informally, we are looking to place multiple 
centers to satisfy a number of customers. Our goal is to minimize the maximal 
distance of a customer to its closest center. For example, in disaster 
management we want to find good locations for camps to provide services to 
multiple cities. More formally, given a set of locations $V$ and number of 
centers to place ($k$) and a distance function $d$ we have to determine a subset 
$S \subseteq V$  with $|S| = k$. The optimization objective $D$ is the minimization of the 
maximal distance of locations to their closest center (see Equation 
\ref{distance}). For this paper the distance function $d$ is based on the 
Euclidean distance in $R^2$. 

\begin{equation}
 \label{distance}
 D = \min \max_{v \in V} \min_{s \in S} d(v,s)
\end{equation}

In this paper we consider only the placement of centers at a customer location 
(Node-Placement). We do not discuss the more general Free-Placement, where 
customers can be placed anywhere. The results and approaches presented in this 
paper can, however, be extended to consider Free-Placement. We focus on 
Node-Placement since it is oftentimes reasonable to rely on the existing 
infrastructure, e.g., given network infrastructure for CDNs or existing roads 
for logistics problems. Especially, for problems with a short planning horizon 
like disaster management or dynamic adaption of CDNs it is impractical to setup 
a new center, that is not at an customer location. When we consider time 
critical applications we need an optimization approach that is fast and robust. 
The following section will briefly introduce a number of optimization 
techniques for the $k$-center problem.

\section{OPTIMIZATION APPROACHES}

Previously, we mentioned the NP-hard nature of the $k$-center problem, which 
makes it a difficult problem to solve. There is no fast exact approach to 
calculate a solution for a given problem instance. Small instances can be solved 
optimally using brute force or branch and bound techniques. Larger scenarios 
cannot be solved practically since the search space grows exponentially. To handle these 
instances a number of heuristic approaches have been proposed to generate 
approximate solutions. Generally, as long as they have polynomial runtime these 
heuristics can only be guaranteed to be 2-approximable, i.e., at worst the 
maximal distance can be twice as large as the theoretical optimum \cite{Kleinberg2005}.  

Dragoon is an 2-approximbale approach that generated very promising results in 
previous experiments \shortcite{hillmann2015c}. We will compare it to the 
following techniques: 2-Approx, MacQueen, Greedy, and a new extension of Greedy 
called Backtrack. A special focus is put on worst case performance of Dragoon 
in comparison to the reference approaches. Dragoon and the other techniques are 
presented in the next few sections. All of them have two things in common, they 
are reasonably fast and they can be implemented to operate in a nearly 
deterministic way without many stochastic effects. This allows us to perform a 
huge number of experiments for different problem instances, without requiring 
experiment repetitions for statistical significant results. 

We have compared Dragoon to other approaches before, that we do not include in 
this paper. Typical metaheuristics like evolutionary algorithms and particle swarm 
optimization are left out because of their stochastic nature. Integer linear 
programming is excluded from the experiments since it requires to much 
calculation time to evaluate large instances effectively. 

\subsection{Dragoon}

The Dragoon algorithm was initially used to optimize Landmark positions to 
improve the quality of IP geolocation \shortcite{hillmann2015a,hillmann2015b}. 
However, it can be applied to any kind of $k$-center problem. Dragoon uses three 
stages for optimization: first an initial reference placement, second center 
placement based on the 2-Approx algorithm (see next section), and finally an 
iterative improvement strategy.

For the initialization, Dragoon places a single virtual center by solving 
the 1-center problem. This means it determines the optimal position for a 
single center. This center will only be used for the first optimization step and 
is removed afterward, consequently, it is not part of the final solution. The 
advantage of this approach is that oftentimes the first center serves more 
customers than centers placed later on. By removing the virtual center we 
generate solutions that are much more balanced with regard to the numbers of 
customers assigned to each center.

In the second stage, Dragoon iteratively places centers using the 2-Approx strategy 
to position all centers. The first placement decision is incorporating the 
virtual center from the initialization step. Afterward, the virtual center is 
removed and no longer specifically considered in further placement decision. 
When the second stage concludes we have a guaranteed 2-approximable solution, 
which is then improved in the final stage.

During the final stage, a local optimization strategy is used to obtain a 
better solution. The procedure iterates over all centers and tries to move them 
to an adjacent node. If such a move leads to an improvement it is accepted, 
otherwise it will be undone. Consequently, the algorithm can only improve the 
initial solution and therefore leads to overall better result. This process is
repeated until no further improvement can be obtained. In the next 
section we will discuss the 2-Approx approach that is used in Dragoon's second stage.

\subsection{2-Approx}

The 2-Approx algorithm discussed in \citeN{hochbaum1985} and is named this way, 
since it reliably provides a 2-approximable solution for the maximum distance ($D$) 
objective. This means, the following is true for all solutions generated by 
2-Approx: 

$$D_{2-Approx} \leq D_{optimal} \cdot 2.$$

The algorithm uses an iterative approach for center placement. The first center 
is placed arbitrarily, either randomly or using some kind of initialization. 
Each successive center is placed at the customer location that has the maximal 
distance to its closest center. This procedure is repeated until all centers 
are 
positioned.

\subsection{MacQueen}

MacQueen is an implementation of a k-means approach \shortcite{macqueen1967}. 
Starting from an initial placement it iteratively optimizes a solution until no 
further improvements are achieved. For each iteration customers are assigned to 
their closest center. Then each center location is changed to optimally support 
the assigned customers. The repositioning is based on the geometrical mean of 
the customer group. Eventually, center positions no longer change and the 
optimization ends. \citeN{selim1984} have shown that k-means approaches are 
guaranteed to converge.

\subsection{Greedy}

We use a straight forward implementation of a greedy approach discussed in 
\shortciteN{jamin2001}. We iteratively place centers one at a time. To place a 
center, 
all possible locations are evaluated and the center is placed at the position 
that maximizes the optimization objective. For node-based placement this means, 
we consider all customer locations during each iteration. With regard to 
free placement, some kind of rasterization is required. Similar to the 2-Approx 
approach, Greedy also guarantees a 2-approximable result.

\subsection{Backtrack}

Backtrack is an extension of the Greedy algorithm we propose. Generally, each 
placement decision in Greedy is only locally optimal. Therefore, the generated 
solution is usually not a global optimum. Backtrack takes the solution generated 
by Greedy and tries to improve it. It only performs changes that lead to an 
improvement, consequently, it only generates solutions that are better or at 
least equally good as the ones generated by Greedy. For this reason, Backtrack 
also guarantees 2-approximable results. For the optimization Backtrack 
individually evaluates all centers and tries to reposition them. This is done 
iteratively for every available center. The optimization terminates either when 
we cannot find a better position for single center after we iterated over all 
of 
them or when a given number of optimization steps is reached. We have to limit 
the maximal number of optimization steps, since we cannot predict how long the 
optimization would run otherwise. In contrast to MacQueen, there is no 
guarantee 
of convergence.

\section{EXPERIMENT SETUP}
\label{experiment-setup}

In our experiments we always consider the relative performance of two 
approaches to solve a given problem instance. Usually, we look at the 
performance of an ``challenger'' algorithm in comparison to a ``challenged'' 
approach (usually Dragoon). To measure the difference between them we simply 
calculate the absolute difference for a given problem instance: 
$$ \Delta D = D_{challenger} - D_{challenged}.$$
Therefore, a negative $\Delta D$ indicates that 
the challenger performed better than the challenged (Dragoon). 

The first part of our experiments was designed to evaluate the average 
performance of the algorithms. We considered six general setups of the $k$-center 
problem (see Table \ref{table:setup}). Instance of these setups were generated 
randomly by placing the customers at random position in a square with $0.0 < 
x,y < 100.0$. We usually selected square numbers for the amount of customers 
and centers, since we can easily map them into the square area for thought 
experiments. For example, if we consider 4 centers that are placed exactly in 
the middle of the 4 sub-quadrants (Free Placement) of our area with side length 
100.0 we know that $D$ can at most be 35.36 ($\frac{1}{4}  \sqrt{100^2 
+100^2}$). For our experiments we evaluated 1000 random instances for each 
comparison. 

\begin{table}[htbp]
 \begin{center}
 \caption{Experiment setups \label{table:setup}}
\begin{tabular}{|l|cccccc|}
\hline
Setup & I & II & III & IV & V & VI \\
\hline
Customers & 10 & 25 & 36 & 49 & 49 & 64\\
Centers & 2 & 4 & 4 & 9 & 4 & 16\\
\hline
\end{tabular}
\end{center}
\end{table}

In the second part of our experiments we tried to identify problem instance 
were the challenger approach would significantly outperform Dragoon. To find 
these instance we employed an evolutionary algorithm that we implemented using the 
Serein framework \cite{uhlig2015}. The evolutionary algorithm used the following setup:

\begin{itemize}
 \item \emph{Problem encoding}: For n customers we used a vector of doubles 
($0.0 < z <1.0$) containing an x and y component for each customer location 
$\vec{v} =(z^x_1,z^y_1, ... ,z^x_n,z^y_n)$. A pair of components $z$ is mapped 
to a location in our square ($100.0^2$).
 \item \emph{Fitness function}: Using the mapped customer locations the centers 
were placed using the challenging and challenged algorithm. Afterward, the 
fitness was calculated as $\Delta D$
 \item \emph{Reproduction operators:} we used single point Gaussian mutation 
with a standard deviation of 0.05 and whole arithmetic recombination with a 
probability of 0.3.
 \item \emph{Selection operators:} we employed tournament selection with 
tournament size 2 and deterministic selection for survival with uncapped elitism.
  \item We used a population size of 20 and run the evolutionary algorithm for 100 
generations.
\end{itemize}

Starting from a random population of scenarios the evolutionary algorithm searches 
for a fitting scenario. In our context fitting means, a scenario that can be 
solved by the challenger approach with a better result than the solution 
provided by Dragoon. The next section will discuss the result of both 
experiments.   

\section{RESULTS}
\label{experiment-results}

The first part of our experiments aimed to confirm the results of our previous 
studies. We expected Dragoon to perform better than the previously considered 
reference approaches, i.e., MacQueen, 2-Approx, and Greedy. At the same time we 
included the new approach Backtrack to evaluate its performance. Looking at the 
results in Table \ref{table:results1}, we can confirm that Dragoon on average returns 
comparably good results. However, Backtrack slightly outperform Dragoon. 

\begin{table}[htbp]
\centering
\caption{Average performance of approaches in comparison to \emph{Dragoon}  
($\Delta D_{avg}$) based on 1000 random problem instances (smaller values are 
better).\label{table:results1}}
\begin{tabular}{|c|d{4.5}d{4.5}d{4.5}d{4.5}|}
\hline
Users / Centers & \multicolumn{1}{c}{MacQueen} & \multicolumn{1}{c}{2-Approx} & 
\multicolumn{1}{c}{Greedy} & \multicolumn{1}{c|}{Backtrack}\\
\hline
 10 / 2  & 4.95 & 13.31 & 4.59 & -0.71 \\
 25 / 4  & 7.64 & 9.09 & 9.5 & 0.70\\
 36 / 4  & 7.84 & 10.13 & 11.02 & -0.59\\ 
 49 / 9  & 6.92 & 3.72 & 6.42 & -1.23\\
 64 / 4  & 7.43 & 12.12 & 12.61 & -0.71\\
 64 / 16 &  6.10 &  1.68 & 2.77 & -1.35\\
\hline
\end{tabular}
\normalsize
\end{table}

\begin{figure}[!htbp]
\centering
\begin{minipage}{.49\textwidth}
  \fbox{\includegraphics[width=\textwidth]{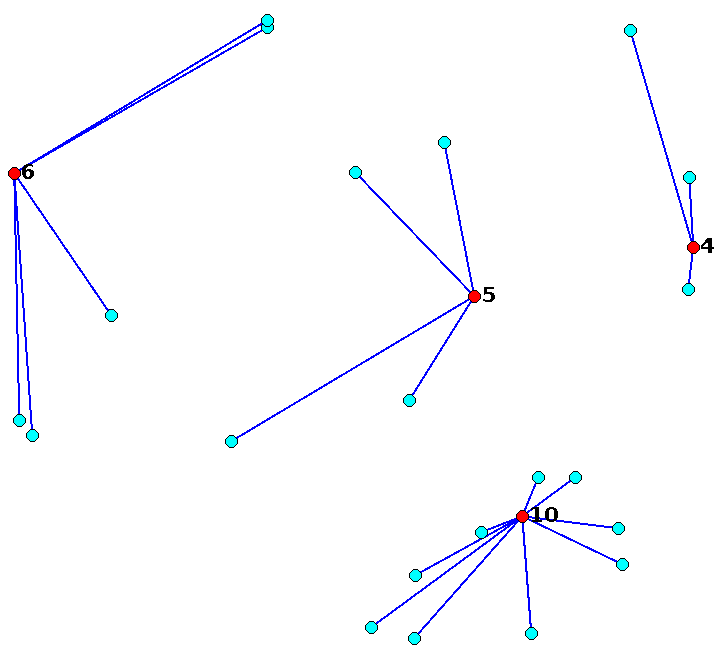}}
  
\end{minipage}%
\hfill
\begin{minipage}{.49\textwidth}
  \fbox{\includegraphics[width=\textwidth]{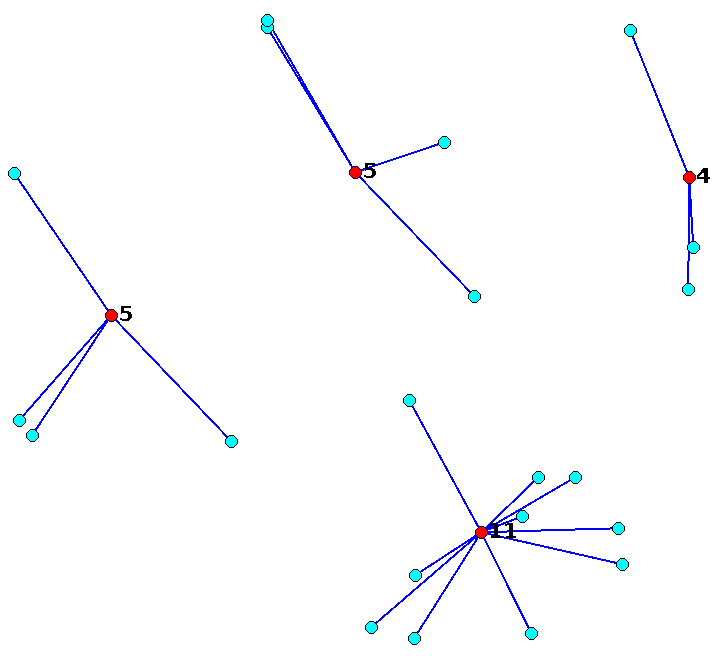}}
\end{minipage}
\caption{For the given scenario Dragoon (left) finds only a suboptimal solution. 
MacQueen (right) determines a much better solution.\label{fig:scenario}}
\end{figure}

The second part of our experiments considered the worst case performance of Dragoon. Table \ref{table:results2} illustrates that we were able to determine problem instance were Dragoon failed in comparison to the reference approach. Conversely, we can say that for each approach there exist scenarios where they can significantly outperform Dragoon. For example, Figure \ref{fig:scenario} shows a scenario where the MacQueen algorithm determines a much better solution than the Dragoon approach. Interestingly Backtrack, our new approach, showed a slightly better average performance than Dragoon.

\begin{table}[htbp]
\centering
\caption{Worst case performance of \emph{Dragoon} in comparison to other 
approaches for a problem instance generated by a GA (smaller $\Delta D$ values  
are better). \label{table:results2}}
\begin{tabular}{|c|d{4.5}d{4.5}d{4.5}d{4.5}|}
\hline
Users / Centers & \multicolumn{1}{c}{MacQueen} & \multicolumn{1}{c}{2-Approx} & 
\multicolumn{1}{c}{Greedy} & \multicolumn{1}{c|}{Backtrack}\\
\hline
 10 / 2  & -26.41 & -29.43 & -31.78 & -23.55 \\
 25 / 4  & -17.34 & -10.54 & -10.81 & -13.39\\
 36 / 4  & -3.23 & -10.84 & -11.96 & -14.85\\ 
 49 / 9  & -8.75 & -5.92 & -4.92 & -8.20\\
 64 / 4  & -7.81 & -4.73 & -0.82 & -10.32\\
 64 / 16 & -5.49 & -5.41 & -4.65 & -6.05\\
\hline
\end{tabular}
\normalsize
\end{table}

\section{DISCUSSION}

Our results illustrate that there are worst case scenarios where Dragoon is 
clearly outperformed by other approaches. In critical application areas this 
can lead to lower quality solutions than initially expected. For these scenarios 
it is important to acknowledge the potential of failure and include appropriate 
fall-back methods. Looking at the scenarios where Dragoon performed worst, we 
noticed that they often had single customers which where located in isolated
positions. However, currently we have no empirical data to statistically support 
this observation. 

It is important to note, that occurrences of low quality solutions are not 
limited to Dragoon. Each of the presented approaches can fail given a certain 
problem instance. In general, nearly every approach can be beaten by all other 
techniques in certain constellations. Table \ref{table:results3} illustrates 
this fact by matching each approach against all the other ones. Here we show 
only one setup, however, the general result was the same across all 
configurations.

\begin{table}[b]
\centering
\caption{Summary of results: Can a certain ``challenger'' approach outperform another ``challenged'' approach in an instance of the $k$-center problem with 49 customers and 4 centers (smaller $\Delta D$ values  are better). 
\label{table:results3}}
\begin{tabular}{|c|d{4.5}d{4.5}d{4.5}d{4.5}d{4.5}|}
\cline{2-6}
\multicolumn{1}{c|}{}&\multicolumn{5}{|c|}{\textbf{Challenged}}\\
\hline
\multicolumn{1}{|c|}{\textbf{Challenger}} & 
\multicolumn{1}{c}{MacQueen}&\multicolumn{1}{c}{Backtrack}&\multicolumn{1}{c}{
TwoApprox}&\multicolumn{1}{c}{Greedy}&\multicolumn{1}{c|}{Dragoon}\\	
\hline
MacQueen	&	&-6.29	&-24.38	&-22.68	&-7.81\\
Backtrack	&-32.68	&	&-24.73	&-26.45	&-10.32\\
TwoApprox	&-31.26	& 0.0	&	&-21.82	&-4.37\\
Greedy		&-21.54	&-0.39	&-15.47	&	&-0.82\\
Dragoon		&-38.53	&-8.52	&-24.64	&-22.12	&	\\	
\hline
\end{tabular}
\normalsize
\end{table}

One exception is the comparison between Greedy and Backtrack, usually Greedy 
should not be able to outperform Backtrack. Backtrack strictly improves a 
solution generated by Greedy, therefore, it should be strictly superior. 
Looking at the results in Table \ref{table:results3}, we see an observation that 
seems to contradict this guarantee, i.e, Greedy outperforms Backtrack with 
$\Delta D = -0.39$). At this point it is important to remember that we selected 
algorithms that are mostly deterministic, nevertheless, there is still some 
randomness included. Specifically, for Greedy if two potential placements lead 
to the same improvement the one that is actually selected is chosen randomly. 

In Table \ref{table:results3}, 2-Approx was not able to outperform Backtrack, but 
in principle it is able to do so. For other setups we have found instances where 
it is better than Backtrack. Nonetheless, the fact that it did not outperform 
Backtrack in the given example indicates that it is quite difficult to find a 
instance were it is superior.

\section{SUMMARY AND OUTLOOK}

We presented a critical view at the Dragoon optimization approach, considering 
its potential worst case behavior. While it returns very good results on 
average, there are instances of the $k$-center problem where it is outperformed 
significantly. The novel Backtrack approach generated very good results and is 
even slightly better than Dragoon. However, as discussed before, it is also 
susceptible to fail for certain scenarios. The general verdict is that all 
approaches shine for certain scenarios and fall off in other cases.

This paper considered the $k$-center problem on a very abstract level. Our future 
work will focus more on practical aspects of the problem. For example, like 
nonlinear distance calculation between customer and its center including 
factors like dynamic traffic, limited availability of certain services depending 
on the center. This includes more intricate assignment strategies that match 
customers and centers based on customer profiles.

With regard to the underlying optimization there are three potential paths to 
follow. The first and simplest approach to guarantee more robustness is to use 
more than one of the presented optimization techniques, to limit the occurrence 
of worst case behavior. Considering time critical application this path might, however, 
be impossible. As a second path, we can add some intelligence to select the 
appropriate approach for a given problem instance. We could either try to train 
a neural network to classify the problem instances or strive to identify 
certain problem properties that favor certain approaches, e.g., avoiding Dragoon 
for problems were certain customers have secluded positions. Finally, we are 
also investigating opportunities to further increase the robustness of the 
approaches on an algorithmic level. Currently we are developing an advanced 
version of Dragoon.

% Please don't exchange the bibliographystyle style
\bibliographystyle{wsc}
% AUTHOR: Include your bib file here
\bibliography{wsc}

\begin{thebibliography}{}

\bibitem[\protect\citeauthoryear{Gonzalez}{Gonzalez}{1985}]{gonzales1985}
Gonzalez, T.~F. 1985.
\newblock ``Clustering to Minimize the Maximum Intercluster Distance''.
\newblock {\em Theoretical Computer Science\/}~38:293 -- 306.


\bibitem[\protect\citeauthoryear{Hillmann, Stiemert, Rodosek, and
  Rose}{Hillmann et~al.}{2015a}]{hillmann2015a}
Hillmann, P., L.~Stiemert, G.~D. Rodosek, and O.~Rose. 2015a, Dec.
\newblock ``Dragoon: Advanced Modelling of IP Geolocation by Use of Latency
  Measurements''.
\newblock In {\em 2015 10th International Conference for Internet Technology
  and Secured Transactions (ICITST)},  438--445.

\bibitem[\protect\citeauthoryear{Hillmann, Stiemert, Rodosek, and
  Rose}{Hillmann et~al.}{2015b}]{hillmann2015b}
Hillmann, P., L.~Stiemert, G.~D. Rodosek, and O.~Rose. 2015b, Nov.
\newblock ``Modelling of IP Geolocation by Use of Latency Measurements''.
\newblock In {\em Network and Service Management (CNSM), 2015 11th
  International Conference on},  173--177.

\bibitem[\protect\citeauthoryear{Hillmann, Uhlig, Rodosek, and Rose}{Hillmann
  et~al.}{2015}]{hillmann2015c}
Hillmann, P., T.~Uhlig, G.~D. Rodosek, and O.~Rose. 2015, Nov.
\newblock ``A Novel Approach to Solve $k$-Center Problems with Geographical
  Placement''.
\newblock In {\em Service Operations and Logistics, and Informatics (SOLI),
  2015 IEEE International Conference on},  31--36.

\bibitem[\protect\citeauthoryear{Hochbaum and Shmoys}{Hochbaum and
  Shmoys}{1985}]{hochbaum1985}
Hochbaum, D.~S., and D.~B. Shmoys. 1985.
\newblock ``A Best Possible Heuristic for the $k$-Center Problem''.
\newblock {\em Mathematics of Operations Research\/}~10:180--184.


\bibitem[\protect\citeauthoryear{Jamin, Jin, Kurc, Raz, and Shavitt}{Jamin
  et~al.}{2001}]{jamin2001}
Jamin, S., C.~Jin, A.~R. Kurc, D.~Raz, and Y.~Shavitt. 2001.
\newblock ``Constrained Mirror Placement on the Internet''.
\newblock In {\em INFOCOM 2001. Twentieth Annual Joint Conference of the IEEE
  Computer and Communications Societies}, Volume~1,  31--40.

\bibitem[\protect\citeauthoryear{Kleinberg and Tardos}{Kleinberg and
  Tardos}{2005}]{Kleinberg2005}
Kleinberg, J., and E.~Tardos. 2005.
\newblock {\em Algorithm Design}.
\newblock Boston, MA, USA: Addison-Wesley Longman Publishing Co., Inc.


\bibitem[\protect\citeauthoryear{MacQueen}{MacQueen}{1967}]{macqueen1967}
MacQueen, J. 1967.
\newblock ``Some Methods for Classification and Analysis of Multivariate
  Observations''.
\newblock In {\em Proceedings of the Fifth Berkeley Symposium on Mathematical
  Statistics and Probability, Volume 1: Statistics},  281--297.

\bibitem[\protect\citeauthoryear{Nguyen, Yosinski, and Clune}{Nguyen
  et~al.}{2014}]{nguyen2014}
Nguyen, A.~M., J.~Yosinski, and J.~Clune. 2014.
\newblock ``Deep Neural Networks Are Easily Fooled: High Confidence Predictions
  for Unrecognizable Images''.
\newblock {\em Computer Vision and Pattern Recognition\/}~abs/1412.1897.


\bibitem[\protect\citeauthoryear{Selim and Ismail}{Selim and
  Ismail}{1984}]{selim1984}
Selim, S.~Z., and M.~A. Ismail. 1984.
\newblock ``$k$-Means-Type Algorithms: A Generalized Convergence Theorem and
  Characterization of Local Optimality''.
\newblock {\em IEEE Transactions on Pattern Analysis and Machine
  Intelligence\/}~6 (1): 81--87.


\bibitem[\protect\citeauthoryear{Uhlig}{Uhlig}{2015}]{uhlig2015}
Uhlig, T. 2015.
\newblock {\em Self-Replicating Individuals}.
\newblock M\"{u}̈nchen: Dr. Hut.


\end{thebibliography}

\section*{AUTHOR BIOGRAPHIES}

\noindent {\bf TOBIAS UHLIG} is a researcher at the Universit\"{a}t der 
Bundeswehr M\"{u}nchen, Germany. He received his M.S. degree in computer science 
from Dresden University of Technology and a Ph.D. degree in computer science 
from the Universit\"{a}t der Bundeswehr M\"{u}nchen. His research interests 
include metaheuristics, simulation based optimization, and process modeling. He 
is a member of the ASIM and IEEE RAS Technical Committee on Semiconductor 
Manufacturing Automation. His email address is \email{tobias.uhlig@unibw.de}.\\

\noindent {\bf PETER HILLMANN} is a Ph.D. student at the Universit\"at der 
Bundeswehr M\"unchen, Germany. He holds a M.Sc. in Information-System-Technology 
from Dresden University of Technology since 2011. His areas of research are 
network and system security with focus on cryptography as well as scheduling and 
optimization problems. His email address is \email{peter.hillmann@unibw.de}.\\

\noindent {\uppercase{\bf OLIVER ROSE}} holds the Chair for Modeling and 
Simulation at the Department of Computer Science of the Universit\"{a}t der 
Bundeswehr M\"{u}nchen, Germany. He received an M.S. degree in applied 
mathematics and a Ph.D. degree in computer science from W\"{u}rzburg University, 
Germany. His research focuses on the operational modeling, analysis and material 
flow control of complex manufacturing facilities, in particular, semiconductor 
factories. He is a member of IEEE, INFORMS Simulation Society, ASIM, and GI. His 
email address is \email{oliver.rose@unibw.de}.\\

\end{document}